\documentclass[twocolumn,showpacs]{revtex4}
\usepackage{amssymb}
\usepackage{graphicx}
\usepackage{dcolumn}
\usepackage{amsmath}
\usepackage{epstopdf}
\usepackage{hyperref}
\begin{document}

\title{Thermo-mechanical sensitivity calibration of nanotorsional magnetometers}

\author{Joseph Losby$^{1,2}$}
\author{Jacob A.J. Burgess$^{1,2}$}
\author{Zhu Diao$^{1,2}$}
\author{David C. Fortin$^{1}$}
\author{Wayne K. Hiebert$^{1,2}$}
\author{Mark R. Freeman$^{1,2}$}

\affiliation{$^{1}$Department of Physics, University of Alberta, Edmonton, Alberta T6G 2G7, Canada}
\affiliation{$^{2}$National Institute for Nanotechnology (NINT), Edmonton, Alberta T6G 2M9, Canada}
\date{Nov 03, 2011}

\begin{abstract}
We report on the fabrication of sensitive nanotorsional resonators, which can be utilized as magnetometers for investigating the magnetization dynamics in small magnetic elements.  The thermo-mechanical noise is calibrated with the resonator displacement in order to determine the ultimate mechanical torque sensitivity of the magnetometer.
\\
\\
\textit{Copyright 2012 American Institute of Physics. This article may be downloaded for personal use only. Any other use requires prior permission of the author and the American Institute of Physics. The following article appeared in the Journal of Applied Physics and may be found at \href{http://link.aip.org/link/?JAP/111/07D305}{http://link.aip.org/link/?JAP/111/07D305}.}

\end{abstract}

\maketitle
The combination of improving nanofabrication and detection techniques, along with technological demand, are providing the means and impetus for further understanding of the micromagnetic structure and dynamics of small magnetic elements.  The exquisite sensitivities of nano- and micromechanical resonators have recently been utilized for detecting quantum-limited displacements \cite{cleland, teufel} and ultra-sensitive atomic-scale mass detection \cite{li, chiu}.  Nanomechanical torque magnetometry allows for direct measurements of single elements \cite{davis,moreland}, offering complementary information to other magnetometry methods, while also allowing for the investigation of details that are often indistinguishable when examining arrays of small elements or larger-scale thin films.  Such events can be correlated to surface inhomogeneities and edge defects of the magnetic element, which play a dominant role on the overall magnetization configuration at smaller size scales.  For understanding of the local energy landscapes of small magnetic elements, it is necessary to further increase the sensitivities of mechanical resonators.

The thermo-mechanical noise sets a fundamental limit for the sensitivity of resonant mechanical measurements.  This Brownian motion is the result of the coupling of the mechanical subsystem to that of a dissipative reservoir (the 'vacuum'), and is generalized as the fluctuation-dissipation theorem \cite{callen, nyquist}.  This coupling causes the resonator to experience position fluctuations at its modes of mechanical resonance.  Following equipartition, the observed mechanical resonance power spectrum must be proportional to the effective temperature of the mechanical mode.  This relationship provides a method for torque sensitivity calibration of torsional resonators.  

Here, we briefly present the fabrication procedure of a doubly-clamped nanomechanical resonator with a permalloy (Ni$_{80}$Fe$_{20}$) disk selectively deposited on a center paddle.  A magnetic driving scheme, where the actuation of the resonator's torsional resonance modes are favored, is described and magnetometry measurements are presented.  Using observations of the thermo-mechanical noise in the un-driven paddle resonators, we offer a calibration method to determine the mechanical torque sensitivity and consequently the magnetic moment sensitivity of the magnetometer, given the strength of the torquing field.

\begin{figure}
\includegraphics[scale=1]{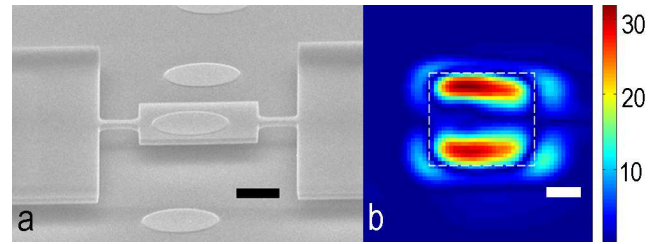}
\caption{(color online) (a) Scanning electron micrograph of a torque magnetometer with a permalloy disk deposited on the paddle. The paddle width is $w_p$=5.8 $\mu$m and the width of the torsion rod is $w_t$=600 nm. The length of one torsion rod is $l$=1.8 $\mu$m and the device thickness is $t$=300 nm.  (b) Spatial raster scan of the lock-in magnitude signal at the torsional resonance frequency of the resonator (top-down view), showing the maximum intensity at the free-edges of the paddle.  The paddle position is indicated by the white dashed line.  The colorbar scale is in mV$_{RMS}$.  Both scale bars are 2 $\mu$m in length.}\label{all}
\end{figure} 

The completed torsional resonator is shown in Fig. 1a. The paddle resonators were fabricated from silicon-on-insulator (SOI) wafers with a 300 nm thick Si device layer and a buried silicon oxide (BOx) layer with a thickness of 1 $\mu$m.  The geometry of an array of paddles was defined via electron beam lithography (EBL) using a negative tone resist (hydrogen silsesquioxane, HSQ).  Following EBL resist development, the unmasked device layer was exposed to a reactive ion etch. A subsequent wet-etching of the BOx layer was done to release the resonators.  The fabricated torsion bar width is 600 nm and the center paddle dimensions are 5.8 $\mu$m x 5.8 $\mu$m.  A commercial silicon nitride membrane (Electron Microscopy Sciences) with a square array of circular holes (12 $\mu$m pitch) served as a deposition shadow mask and was aligned above the device chip using a vacuum-tweezer micro-manipulator.  Collimated electron beam deposition, under 10$^{-10}$ mbar vacuum, of permalloy resulted in the placement of magnetic disks on the paddles of the resonators.  The thickness of the disks are approximately 35 nm with a diameter of 3.2 $\mu$m, measured through scanning electron microscopy.  

The torque exerted on the magnetization of the permalloy disk by an external field acts as a mechanical load, resulting in the deflection of the resonator.  In order to drive the torsional mode of the resonator, a bias field (provided by a permanent magnet housed on a motorized stepper rail) was oriented in-plane with the disk at a direction to induce a non-zero magnetization component orthogonal to the resonator length.  A copper wire coil placed above the devices provided a small time-varying, out-of-plane 'dither' field.  A high frequency lock-in amplifier (Zurich Instruments Model HF2) provided the sinusoidal drive/reference signal in the range of the resonator's torsional resonance mode.  The vertical displacement of the resonator was detected by taking advantage of optical interferometry.  The SOI architecture allows for a controlled optical cavity spacing between the resonator and handle layer, providing the interferometric contrast for sensitive displacement measurements.  The laser ($\lambda=$632 nm) was focused through an objective lens and the modulation of the reflected intensity was detected using a low-noise, RF photodetector (New Focus Model 1801).  The measurements were taken in vacuum under a pressure of $\sim$10$^{-6}$ mbar.  A spatial  raster scan of the interferometric signal is shown in Fig 1b., confirming the driving of the torsional resonance mode. 

\begin{figure}
\includegraphics[scale=1]{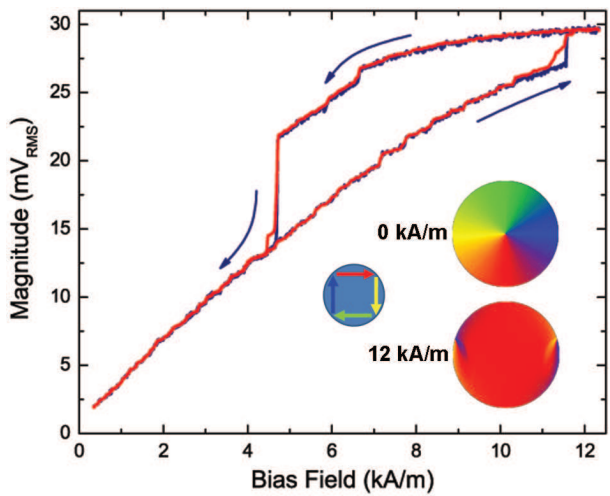}
\caption{(color online) Magnetic vortex hysteresis loop in the permalloy disk.  The dark blue trace is a single bias field sweep and the red trace is averaged for a $\sqrt{20}$ factor of noise reduction.  The drive field was 150 A/m and the lock-in amplifier time constant was 120 ms.  The results for the micromagnetic simulations of the disk are shown as colormaps, where the in-plane magnetization vector directions are represented by the color wheel.} \label{all}
\end{figure} 

The acquired lock-in magnitude signal of the driven magnetometer is shown in Fig. 2 as a unipolar bias field was swept.  A magnetic vortex hysteresis loop was captured, where the sharp transitions at the lower and higher fields represent the vortex creation and annihilation fields, respectively.  The magnetization states at the different field regions are supported by Landau-Lifshitz-Gilbert based micromagnetic simulations of the hysteresis, shown in Fig. 2 at 0 and 12 kA/m.  The dark blue trace is a single bias field sweep, while the averaged signal of 20 sweeps is shown in red. The reproducibility of the smaller-scale, Barkhausen-like transitions throughout the measurement, due to inhomogeneities within the permalloy disk, are seen in the single trace with sufficient signal-to-noise.  These devices can possibly be implemented to probe and map the potential energy landscapes of small magnetic elements.

With the driving field off, the thermo-mechanical power spectral density, $S_{V}(f)$, around the torsional resonance was acquired by sweeping the lock-in frequency, Fig. 3.  The angular displacement spectral density of the thermal vibrations is given by,
\begin{equation}
S_\theta(f)=\frac{2k_BTf_0^{3}}{\pi k_{tor}Q}\frac{1}{(f_0^{2}-f^{2})^{2}+(\frac{f_0f}{Q})^{2}},
\end{equation}
where $T$ is the effective temperature of the resonator, $f_{0}$ is the center frequency, and $Q$ is the mechanical quality factor \cite{bunch}. $\theta$ is the mean angular displacement of the paddle from its equilibrium position. $k_{tor}$ is the torsional spring constant and is, for the resonator geometry, given by,
\begin{equation}
{k}_{tor}=\frac{4G{t}^{3}(0.33{w}_{t}-0.21t)(0.33{w}_{p}-0.21t)}{0.33(2l{w}_{p}+{w}_{p}{w}_{t})-0.21t(2l+{w}_{p})}, 
\end{equation}
where $G$ is the shear modulus of the resonator material, $t$ is the thickness, $w_{t}$ is the width of the torsion rod, $w_{p}$ is the width of the paddle, and $l$ is the length of a torsion rod \cite{lobontiu}.  The equipartition theorem allows for the calibration of the power spectral density with the displacement spectral density,
\begin{equation}
\left \langle \theta^{2} \right \rangle =\frac{k_BT}{\kappa_{tor}}=\int_{0}^{\infty }S_\theta(f) df.
\end{equation}
Solving Eq. (3) at $T=295$ K using the calculated spring constant $k_{tor} =2.35 \times {10}^{-10}$ N m/rad. results in $\left \langle \theta \right \rangle=4.16$ $\mu$rad., which equates to a mean edge displacement of the paddle from equilibrium as $ \sqrt{ \left \langle x^{2} \right \rangle}=11.8$ pm.  Solving Eq. (1) with (3) and assuming a Lorentzian form for the thermo-mechanical noise allows for the calibration of the theoretical displacement spectral density distribution on resonance, $S_x(f_0)=2Q\left \langle x^2 \right \rangle/ \pi f_0  = 0.093$pm$^2/$Hz, to the power spectral density, $S_v(f_0)$, as shown in Fig. 3.

\begin{figure}
\includegraphics[scale=1]{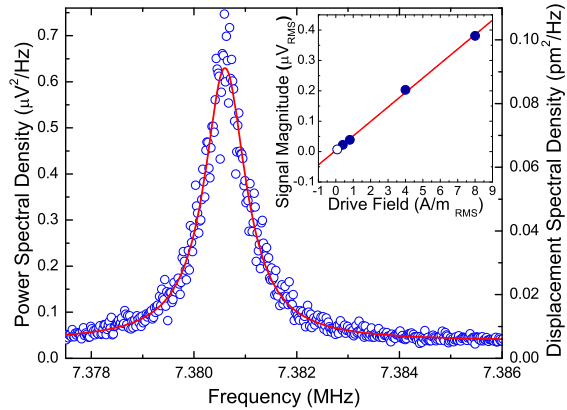}
\caption{(color online) Frequency spectrum of the thermo-mechanical power spectral density, $S_{v}$, calibrated with the displacement spectral density, $S_x $.  The scan is averaged 16x and the noise-equivalent bandwidth is 9.88 Hz.  The solid red trace is a Lorentzian fit, where the values $f_{0}$=7.3806 MHz and $Q$=7775 were extracted.  Inset: calibration of the signal magnitude to the driving field.  The hollow circle is magnitude of the thermo-mechanical limit superimposed onto the linear fit, corresponding to a thermo-mechanically limited drive of 0.12 A/m.}\label{all}
\end{figure}

The responsivity and displacement sensitivity of the magnetometer can be calculated through the determination of the scaling factor, $\alpha$, between $S_x(f)$ and $S_v(f)$ \cite{hiebert}.  That is, $S_v(f)=S_v^{white}+\alpha S_x(f)$, where $S_v^{white}$ is the white noise floor (the sum of the optical shot noise of the laser and the photodetector dark current).  On resonance ($f = f_0$), 0.628$\mu$V$^2 $Hz$^{-1}=$ 0.036$\mu$V$^2 $Hz$^{-1}+\alpha $0.093 pm$^2$Hz$^{-1} $$\rightarrow$ $\alpha=6.32(\mu$V/pm)$^2$.  The responsivity is defined as $R=\alpha^{1/2}=2.5\mu$V/pm.  Using the scaling factor to convert $S_v^{white}$ to the displacement spectral density noise floor, $S_x^{white}=S_v^{white}/\alpha$=0.0058pm$^2$/Hz, the displacement sensitivity of the cantilever off-resonance can be found, $(S_x^{white})^{1/2}$ = 75 fm/Hz$^{1/2}$.  The torque sensitivity away from resonance, upon conversion to an angular displacement, $\tau=\kappa_{tor} \theta$, is 6.2 aNm/Hz$^{1/2}$.  At mechanical resonance this value increases approximately by a factor of $\sqrt{Q}$ \cite{cleland2}, which gives our magnetometer a torque sensitivity of $\sim$ 70 zNm/Hz$^{1/2}$.  

We define the figure-of-merit ($FOM$) as the scaling factor of the magnetic moment sensitivity of a thermally-limited magnetometer with driving field.  The calibration of the linear signal magnitude with the torquing field is shown in the inset of Fig. 3.  The thermo-mechanical limit, represented by the hollow circle superimposed onto the linear fit, corresponds to an a.c. drive field of $H_{tm}=0.12 A/m$.  With the conservative assumption that all of the magnetic moments contribute to the drive at this field, the $FOM$ is then $M_SV \times H_{tm} = \sim 2.5 \times 10^9 \mu_B (A/m)$, where $M_S=800 kA/m$ is the saturation magnetization of permalloy and $V$ is the volume of the disk.  For our measurements (Fig. 2) the driving field was $150 A/m$, leading to a moment sensitivity of $FOM / 150 (A/m) = \sim1.5 \times 10^7 \mu_B$.  A conversion to a torque yields $\vec{\tau}=\vec{m} \times \mu_0 \vec{H}=\sim 26.2$ $zNm$, which is on the order of our mechanical torque sensitivity.  The dynamic range of the moment sensitivities ranges from $\sim2 \times 10^{10} \mu_B$ at the thermo-mechanical limit to $\sim4 \times 10^6 \mu_B$ at the maximum driving field ($600 A/m$) of our instrumentation.  This is approximately an order-of-magnitude improvement in sensitivity to comparable torque magnetometry studies at room temperature using low applied fields \cite{moreland, davis2, moreland2}.   

In this study the fabrication of sensitive nanotorsional magnetometers for the study of single magnetic elements was described, along with magnetometry results of a permalloy disk with a ground-state magnetic vortex.  The post-fabrication deposition of the magnetic material provides the added benefit of the magnetic disk remaining relatively 'clean' (that is, unaffected by acid or plasma etching processes that occur during the resonator fabrication).  The resonators also provide a sensitive platform for the study of various deposited magnetic materials.  We have also demonstrated the capability of calibrating the measured thermo-mechanical noise with the theoretical resonator displacement to determine the ultimate torque sensitivity of the resonators.  Through the tuning of the resonator geometry and transduction scheme it is possible to further enhance the sensitivities of these devices. 

\begin{acknowledgements}
The authors would like to acknowledge support by NSERC, CIFAR, CRC, Alberta Innovates, and NINT.  The devices were partially fabricated at the University of Alberta NanoFab and SEM imaging was performed at the NINT electron microscopy facility.  
\end{acknowledgements}


\begin{thebibliography}{99}

\bibitem{cleland} A.D. O'Connell, M. Hofheinz, M. Ansmann et al., Nature \textbf{464}, 697 (2010).
\bibitem{teufel} J. D. Teufel, T. Donner, Dale Li, J. W. Harlow et al., Nature \textbf{475}, 359 (2011).
\bibitem{li} M. Li, W. H. P. Pernice, and H. X. Tang, Nat. Nanotechnol. \textbf{4}, 377 (2009).
\bibitem{chiu} Y. Chiu, P. Hung, H. W. C. Postma, and M. Bockrath, Nano Lett. \textbf{8}, 4342 (2008).
\bibitem{davis} J.P. Davis, D. Vick, J.A.J. Burgess, D.C. Fortin et al., New J. Phys. \textbf{12} 093033 (2010).
\bibitem{moreland} M.D. Chabot, J.M. Moreland, L. Gao et al., J. Micromech. Microeng. \textbf{14} 1118 (2005).
\bibitem{callen} H.B. Callen and T.A. Welton, Phys. Rev. \textbf{83}, 34 (1951).
\bibitem{nyquist} H. Nyquist, Phys. Rev. \textbf{32}, 110 (1928).
\bibitem{bunch} J. S. Bunch, A. M. van der Zande, S. S. Verbridge et al., Science \textbf{315}, 490 (2007).  $Supporting$ $Online$ $Material$.
\bibitem{lobontiu} N. Lobontiu, B. Ilic, E. Garcia et al., Rev. Sci. Inst. \textbf{77}, 073301 (2006).
\bibitem{hiebert} W.K. Hiebert, D. Vick, V. Sauer, and M.R. Freeman, J Micromech. Microeng. \textbf{20} 115038 (2010).
\bibitem{cleland2} A. N. Cleland, \textit{Foundations of Nanomechanics: from Solid-State Theory to Device Applications.} (Springer, Berlin 2003).
\bibitem{davis2} J.P. Davis, D. Vick, D.C. Fortin, J.A.J. Burgess et al., Appl. Phys. Lett. \textbf{96}, 072513 (2010).
\bibitem{moreland2}  J. Moreland, J. Phys. D: Appl. Phys. \textbf{36}, R39 (2003). 


\end{thebibliography}
\end{document}